\documentclass[5p]{elsarticle}

\usepackage{color}
\usepackage{psfrag}
\usepackage{amsmath,amssymb}
\usepackage{graphicx}
\usepackage{natbib}

\biboptions{authoryear}

\begin{document}

\begin{frontmatter}

\title{Mathematical Modelling of Allergy and Specific Immunotherapy:\\Th1-Th2-Treg Interactions}

	\author[itp]{Fridolin Gro\ss\corref{cor2}\fnref{fn1}}
	\ead{fridolin.gross@ifom-ieo-campus.it}

	\author[ici]{Gerhard Metzner}

	\author[itp]{Ulrich Behn\corref{cor1}}
	\ead{ulrich.behn@itp.uni-leipzig.de}

	\cortext[cor1]{Corresponding author. Tel.: \mbox{+49 341 97 32434.}}
	\cortext[cor2]{Principal corresponding author. Tel.: \mbox{+39 02 9437 5102.}}
	\fntext[fn1]{Present address: Campus IFOM-IEO, Via Adamello 16, 20139 Milan, Italy.}

	\address[itp]{Institute for Theoretical Physics, P.O.B. 100 920, D-04009 Leipzig, Germany}
	\address[ici]{Institute for Clinical Immunology, Johannisallee 30, D-04103 Leipzig, Germany}

\begin{abstract}
	Regulatory T cells (Treg) have recently been identified as playing a central role in allergy and during allergen-specific immunotherapy. We have extended our previous mathematical model describing the nonlinear dynamics of Th1-Th2 regulation by including Treg cells and their major cytokines. We hypothesize that immunotherapy mainly acts on the T cell level and that the decisive process can be regarded as a dynamical phenomenon. The model consists of nonlinear differential equations which describe the proliferation and mutual suppression of different T cell subsets. The old version of the model was based upon the Th1-Th2  paradigm and is successful in describing the ``Th1-Th2 switch" which was considered the decisive event during specific immunotherapy. In recent years, however, the Th1-Th2 paradigm has been questioned and therefore, we have investigated a modified model in order to account for the influence of a regulatory T cell type. We examined the extended model by means of numerical simulations and analytical methods. As the modified model is more complex, we had to develop new methods to portray its characteristics. The concept of stable manifolds of fixed points of a stroboscobic map turned out to be especially important. We found that when including regulatory T cells, our model can describe the events in allergen-specific immunotherapy more accurately. Our results suggest that the decisive effect of immunotherapy, the increased proliferation of Treg and suppression of Th2 cells, crucially depends on the administration of high dose injections right before the maintenance phase sets in. Empirical protocols could therefore be improved by optimizing this step of therapy. 
\end{abstract}


\end{frontmatter}

\section{Introduction}
	T helper cells play a significant role in immune responses to allergic substances. There are several subtypes of T helper cells which differ in function according to their cytokine profiles. Specific Th2 cells are mainly responsible for allergic reactions as they can activate the production of IgE antibodies by means of which the well known allergic symptoms are provoked.  The ``Th1-Th2 paradigm" that has guided immunologists since the late 1980s states that the type of immune response depends on which of the two populations prevails in the concurrence of Th1 and Th2 helper cells. For the case of allergy this entails that there are populations of allergen-specific T helper cells in both allergic and healthy individuals. Yet, in the latter an allergic response is prevented by the predominance of Th1 cells \citep{romagnani1997th1}. The ``hygiene hypothesis", that claims that  a hygienic childhood environment increases the risk of allergic diseases, can be explained in this framework as follows: Due to the reduced exposure to bacterial and viral antigens the Th1 cells are only insufficiently stimulated and therefore cannot prevent the Th2 cells from dominating after exposure to an allergen \citep{yazdanbakhsh2002allergy}.

In recent years doubts have been raised about the persuasiveness of this explanation.  Some studies show, for instance, that populations with high rates of helminth infections are equally protected from allergic diseases, even though these infections induce strong Th2-mediated immune reactions. On the other hand, a considerable increase in the frequency of type 1 diabetes and other autoimmune diseases, which turn out to be mediated by Th1 cells, has been observed \citep{wills2001germless}. Thus, there seems to be yet another mechanism of regulation which is able to prevent the development of unwanted immune responses in healthy individuals and whose malfunction can lead to either allergic or autoimmune disease. A modified version of the hygiene hypothesis known as ``counter-regulation hypothesis" has been suggested according to which all kinds of infections can possibly prevent the development of allergic disorders by inducing the proliferation of regulatory T cells (Treg) \citep{murphy2007janeway,sakaguchi2000regulatory}.

Different types of regulatory T cells have been identified. The type of Treg cells which seems to be important in the context of allergic diseases is the so-called \emph{induced} regulatory T cell (Tr1). Cells of that kind produce cytokines such as IL10 and TGF-$\beta$ which can suppress both Th1 and Th2 mediated immune responses and they differentiate from naive T cells just as the other subsets \citep{battaglia2006tr1,taylor2006mechanisms}.

Allergen-specific immunotherapy (also known as desensitization therapy) consists of repeated injections of allergen or allergen peptides and aims at inducing a state of tolerance in the allergic individual. Even though specific immunotherapy has been carried out for more than one century now, the underlying mechanisms remain poorly understood. Within the framework of the Th1-Th2 paradigm immunologists assumed that in the course of immunotherapy  the Th2 mediated reaction is ``switched" to a Th1 dominated response \citep{murphy2007janeway}.  More recent studies indicate, however, that the therapeutic effect is mainly caused by an increase in the population of allergen-specific regulatory T cells \citep{akdis1998role,akdis2007mechanisms}.

The therapy is performed in practice by starting with very small, innocuous injections which are subsequently increased until  a maximum dose is reached. After that, during the \textit{maintenance phase}, this dose is administered once every four weeks over a period of 3-5 years. There exist different protocols for the initial part of the treatment which differ in the period of time in which they reach the maximum dose. In conventional therapies it takes about two months, while in so called ``rush protocols" the maintenance dose is reached after only one week \citep{bousquet1998allergen}.

If we assume that the therapeutic effect of immunotherapy is mainly due to a change of T cell equilibrium which involves only a small number of cell types, it should be possible to capture it within a mathematical model. A model using nonlinear differential equations describing the dynamics of Th1-Th2-Interactions has been introduced in \cite{behn2000modeling} and was further investigated in \cite{richter2002mathematical} and \cite{vogel2006th1}. On the following pages we will present an extended version of the model that takes into account the influence of the population of allergen-specific regulatory T cells. After motivating the set of equations that defines our model and explaining the occurring parameters, we will investigate the simple case of periodic injections. By making use of the stroboscopic map, we will already be able to anticipate the qualitative features of realistic therapies. Finally, we will show that we can simulate different therapy protocols provided that the initial conditions are chosen in the right way.
	
\section{The Model}
\begin{figure}[h!]
\centering
\psfrag{IgE}{IgE,}
\psfrag{Allergie}{\scriptsize{Allergy}}
\psfrag{T1}{$T_1$}
\psfrag{T2}{$T_2$}
\psfrag{N}{$N$}
\psfrag{APC}{APC}
\psfrag{IFN}{\emph{IF}}
\psfrag{IL}{\emph{IL}}
\psfrag{Tr}{$T_r$}
\includegraphics[width=\columnwidth]{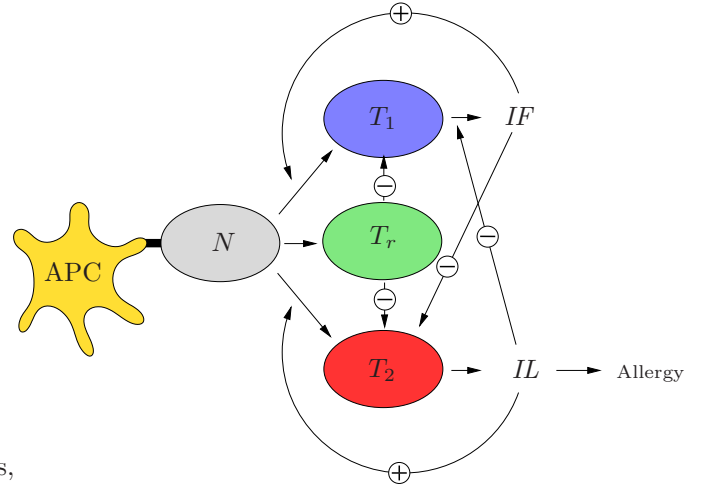}
\caption{Simplified scheme of T cell interaction in response to allergen encounter. Allergen presented by an antigen-presenting cells (APC) activates naive T helper cells ($N$) which leads to their subsequent differentiation to either Th1, Th2, or Treg cells ($T_1$, $T_2$, and $T_r$ respectively). Th1 and Th2 cells suppress each other and support their own proliferation respectively via their cytokines (\emph{IF}, \emph{IL}). Treg cells suppress Th1 and Th2 cells. Dominance of Th2 cells leads to allergic reaction.
\label{fig:scheme}}
\end{figure}

	Our model consists of a set of nonlinear differential equations describing the temporal behavior of five variables: the concentrations of Th1, Th2 and Tr1 cells ($T_1$, $T_2$, $T_r$ respectively), the concentration of naive T helper cells ($N$), and the concentration of allergen ($A$) presented by antigen-presenting cells. Figure \ref{fig:scheme} shows a simplified scheme of the T cell interactions that are incorporated into the model. 
Following an injection, allergen is taken up by an antigen-presenting cell (APC) and presented to naive T helper cells. Upon activation, these naive cells can differentiate into Th1, Th2 or Treg cells. Via their cytokines (\emph{IF}, \emph{IL} respectively) activated cells can exert autocrine action on their own population and suppress proliferation of the other. Th1 and Th2 cells suppress each other respectively, whereas Treg cells suppress both Th1 and Th2 cells while they themselves are not suppressed. The asymmetric way in which the populations of Th1 and Th2 cells interact is adopted from the previous version of the model \citep{behn2000modeling}.  Our new attempt to describe the concurrence of T cells leads to the following set of equations:

\begin{align}
\begin{split}
\dot N=&-N+\alpha-NA\left(\frac{T_1}{1+\mu_2 T_2}+c\right)\\&-\phi NA(T_2+c)-\chi NA(T_r+c)
\end{split}\label{eq1}\,,\\
\dot T_1=&-T_1+\frac{\upsilon NA}{1+\mu_r T_r}\left(\frac{T_1}{1+\mu_2 T_2}+c\right)\label{eq2}\,,\\
\dot T_2=&-T_2+\phi\frac{\upsilon NA}{1+\mu_r T_r} \left(\frac{T_2+c}{1+\mu_1\frac{T_1}{1+\mu_2 T_2}}\right)\label{eq3}\,,\\
\dot T_r=&-T_r+\chi\upsilon NA\left(T_r+c\right)\label{eq4}\,,\\
\dot A=&-A(T_1+T_2+T_r)\label{eq5}\,.
\end{align}
\vspace*{0.5cm}

We shall now explain the form of the equations as well as the occurring parameters. Looking at equations \eqref{eq1}--\eqref{eq3}, we find that the specific T cell populations only grow to substantial sizes if allergen is presented. In the absence of such a stimulus most cells die off. All T cells (including the naive cells) are assumed to have the same half life and consequently all populations decay at the same rate. The system is already rescaled to dimensionless units, in particular the time is measured in units of the half life of T cells. Naive cells are produced at a constant rate $\alpha$, whereas the generation of Th1, Th2, and Treg cells is proportional to the concentration of naive cells, the concentration of presented allergen, as well as to the concentration of their respective cytokines (autocrine stimulation). As the cytokines are degraded fast compared to the half life of cells, the concentration of cytokines produced by a T cell subpopulation can be regarded as proportional to the size of that population itself. For that reason, the cytokines do not explicitly appear in the equations. The parameter $c$ accounts for a small background of cytokines arising from other processes of the immune system. It is assumed to be equal for the three subsets of differentiated helper cells and to be constant over time. Its mathematical role consists in initially driving the system away from the trivial state, where all T cell concentrations are zero. Suppression is modeled by factors of the form $1/(1+x)$ where $x$ stands for the concentration of cytokines produced by the suppressing population. In these factors the small cytokine background is neglected. Finally, equation \eqref{eq5} states that the presented allergen is degraded proportionally to the total concentration of specific T cells. 

The parameter $\upsilon$ determines how many differentiated T cells arise from one naive cell, $\phi$ and $\chi$ account for differences in the autocrine action of the three subsets. The strength of suppression is regulated by the parameters $\mu_1$, $\mu_2$, and $\mu_r$ respectively.

A more detailed derivation for the case of the Th1-Th2 model can be found in \cite{richter2002mathematical}. Furthermore, it is argued there that the parameters $\phi$, $\mu_1$, and $\mu_2$, which already occur in the old version of the model, have to satisfy the conditions $\phi\gtrsim 1$ and $\mu_1>\mu_2$. In the next step we will try to find analogous conditions for the choice of $\chi$ and $\mu_r$. To this end, we will turn our attention to the development in time of the ratios $T_1/T_r$ and $T_2/T_r$. To keep the following calculations simple, we use the approximation $c\approx0$. It can be shown that the conclusions drawn also hold for the case of small but nonvanishing $c$. 

It follows from \eqref{eq2} and \eqref{eq4} that
\begin{align}
\frac{d}{dt}\frac{T_1}{T_r}=\upsilon NA\left(\frac{1}{(1+\mu_r T_r)(1+\mu_2 T_2)}-\chi\right)\,.
\end{align}
Setting this expression equal to zero yields
\begin{align}
T_r=\frac{1}{\mu_r}\left(\frac{1}{\chi(1+\mu_2 T_2)}-1\right)\,.
\label{eqT1/Tr}
\end{align}
In the same way, for the case of $T_2/T_r$ we find
\begin{align}
T_r=\frac{1}{\mu_r}\left(\frac{\phi}{\chi\left(1+\mu_1 \frac{T_1}{1+\mu_2 T_2}\right)}-1\right)\,.
\label{eqT2/Tr}
\end{align}
Equations \eqref{eqT1/Tr} and \eqref{eqT2/Tr} can only be satisfied for positive cell concentrations if we set $\chi<1$ and $\chi<\phi$. Otherwise the Treg cells will always dominate over the other two subsets which makes it impossible to simulate any allergic reaction at all.
Provided that $\chi<1<\phi$, we find that above a threshold given by
\begin{equation}
T^{th}_r=\frac{1}{\mu_r}\left(\frac{\phi}{\chi}-1\right)
\label{eq:param}
\end{equation}
the Treg cells have a higher growth rate than the two other populations. This threshold is independent of the concentrations of Th1 and Th2 cells. If its value is set too high, the Tregs will never be able to compete and there will be no successful therapy. A very low threshold on the other hand will make them too dominant. Therefore, the relation given by \eqref{eq:param} can lead us to a reasonable choice of $\mu_r$. For numerical simulations, we will always use $\chi=0.8$ and $\mu_r=0.25$. From \cite{richter2002mathematical} we adopt the choice of the remaining parameters:
$\alpha=10$, $\upsilon=8$, $\chi=1.02$, $\mu_1=0.2$, $\mu_2=0.1$, and $c=10^{-4}$.

Equations \eqref{eq1}-\eqref{eq5} constitute an autonomous dynamical system, but this only holds because we have not yet considered how the allergen is taken up by the organism. In immunotherapy the allergen enters the body via subcutaneous injections. In our model an injection of allergen at a given time $t$ is modeled by changing the allergen concentration instantaneously from $A(t)$ to $A(t)+D$, where $D>0$ specifies the dose administered. After an injection the three T cell populations expand by several orders of magnitude. However, after a short time (compared to the half life of T cells) the allergen has been degraded completely ($A\approx0$) and the populations will not grow any longer. It follows from equations \eqref{eq2}-\eqref{eq4} that subsequently their concentrations will drop exponentially. As the half life is the same for Th1, Th2, and Treg cells, the ratios of concentrations will by then have reached a constant value. 

According to \cite{akdis2004immune}, it is in particular the balance of Th2 and Treg cells that is decisive as to whether there will be an allergic reaction or not. In our model we can directly compare the initial value of the ratio $T_2/T_r$ to the constant value that is reached after an allergen encounter. We will therefore call an immune response to a given dose $D$ \textit{allergic} if this ratio has increased compared to its initial value, or, mathematically speaking, if
\begin{align}
\lim_{t\rightarrow\infty}\frac{T_2(t)}{T_r(t)}>\frac{T_2^0}{T_r^0}\,.
\label{eqallergic}
\end{align} 
Starting from the assumption that the naive cells are in their stationary state ($N=\alpha$) before the allergen encounter and that there is no allergen left from previous encounters ($A=0$), the type of reaction only depends on the initial conditions $T_1^0$, $T_2^0$, $T_r^0$ and on the allergen dose $D$. 

Of course, this description of allergen administration is highly idealized and it only approximates the case when allergen is taken up in an injection-like fashion, which applies for example to insect stings. Allergic reactions to pollen or house dust mite, however,  are more complicated as the allergen is taken up continuously over time.
	
\section{Fixed Points and Stable Manifolds}
	Our next step will be to describe specific immunotherapy, that is, administration of repeated injections. The mathematically simplest case is that of periodic injections, which means giving the same dose $D$ repeatedly at times $t_0+n\cdot\tau,\,n=0,1,2,3,\ldots$. The case of periodic injections corresponds to the maintenance phase of allergen specific immunotherapy. To investigate it in more detail we will use the \emph{stroboscopic map}. This concept has been proposed in \cite{vogel2006th1} and it has proven to be an important tool because it considerably reduces the complexity of the system. 

We denote by $\theta(\mathbf T^0,\mathbf A^0;t)$ the solution of the above 
system at time $t$ for initial conditions 
\begin{align}
(\mathbf T^0,\mathbf A^0)=(T_1^0, T_2^0, T_r^0, N^0, A^0 )\,,
\end{align}
furthermore, let $\theta_{\mathbf T}(\mathbf T^0,\mathbf A^0;t)$ be the projection of this solution on the three-dimensional subspace of T cell concentrations. Elements of this space are vectors of the form $\mathbf T=(T_1,T_2,T_r)^T$.
The stroboscopic map $\mathbf{S}_{\tau,D}(\mathbf T)$ for period $\tau$ and allergen dose $D$ is then defined as
\begin{align}
\mathbf{S}_{\tau,D}(\mathbf T)=\theta_{\mathbf T}(\mathbf T,\mathbf A_D;\tau)\,,
\label{strobo}
\end{align}
where $\mathbf A_D=(\alpha,D)$.
Equation \eqref{strobo} maps a vector $\mathbf T$ of cell concentrations on the state the system will be in at time $t=t_0+\tau$ if an injection of dose $D$ is given at time $t=t_0$ where $\mathbf T$ is taken as initial condition. If $\tau$ is not too small, then again at time $t_0+\tau$ we will have $N\approx\alpha$ and $A\approx0$. This means that applying the stroboscopic map repeatedly will be a good approximation for describing a periodic therapy.

\begin{figure}[h!]
\centering
\psfrag{ 1e-4}{\footnotesize$10^{-4}$}
\psfrag{ 0.01}{\footnotesize$0.01$}
\psfrag{ 1}{\footnotesize$1$}
\psfrag{ 100}{\footnotesize$100$}
\psfrag{T1}{\footnotesize$T_1$}
\psfrag{T2}{\footnotesize$T_2$}
\psfrag{Tr}{\footnotesize$T_r$}
\includegraphics[width=\columnwidth]{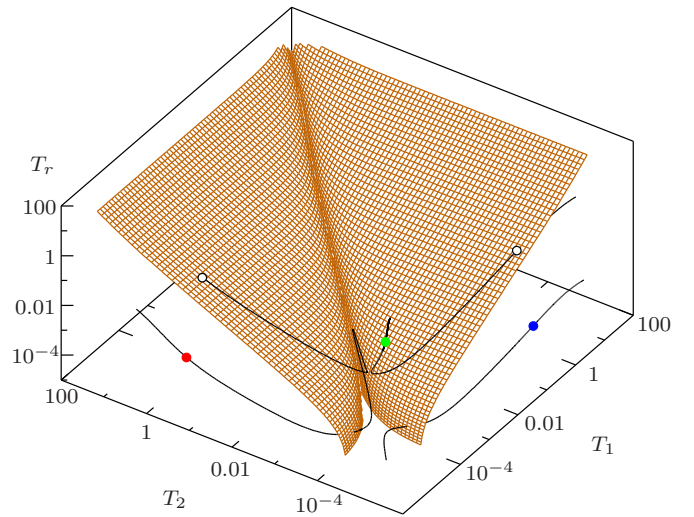}
\caption{Fixed points and stable manifolds. The set of fixed points of the stroboscopic map \eqref{strobo}, for $D=1$ and varying period $\tau$, which is made up of three different branches is displayed. As an example the stable (\textcolor{red}{$\bullet$}\textcolor{green}{$\bullet$}\textcolor{blue}{$\bullet$}) and unstable ($\circ$) fixed points corresponding to $\tau=4$ are shown along with the stable manifolds of the unstable fixed points (brown surfaces). T cell concentrations are represented logarithmically. }
\label{fig2}
\end{figure}

We will now investigate the long term behavior of the system when such a periodic therapy is applied. It follows from what has been said above that we can attack this question by simply looking at repeated applications of the stroboscopic map. If only one single injection is given, the system will eventually reach the trivial state $(\alpha,0,0,0,0)$. However, the perturbations exerted by the repeated injections can create periodic orbits in which the concentration of each cell subpopulation keeps oscillating in a uniform fashion always reaching the same peak value.  These orbits correspond to fixed points of the stroboscopic map. Numerical simulations show that for a given period $\tau$ the stroboscopic map has up to three stable fixed points and several unstable fixed points. If we look at stroboscopic maps for different periods, we find that their fixed points lie on continuous lines. In figure \ref{fig2} these are displayed as branches of stable and unstable fixed points. Bifurcations occur at certain critical periods which means that the number of fixed points can change if $\tau$ is changed. 

Applying the stroboscopic map $\mathbf{S}_{\tau,D}$ repeatedly will drive the system to one of its stable fixed points. In general we find three such fixed points, consequently there are three possible outcomes for the corresponding periodic therapy. In each of the stable fixed points the concentration of one of the T cell subsets peaks high while the concentrations of the two others remain much lower. Administration of periodic injections will therefore always result in one cell type eventually dominating the two others. Which one of the three subsets will in the end be successful crucially depends on the initial state given by a vector $\left( T_1^0,T_2^0,T_3^0\right)\in\mathbb{R}^3_+$. We can therefore subdivide the space of T cell concentrations into three regions, each being the set of all initial vectors leading to the same therapeutic result. These regions are just the domains of attraction of the stable fixed points of the stroboscopic map. The boundaries between the domains of attraction are constituted by the stable manifolds of the unstable fixed points.  In addition to showing the branches of fixed points for varying period $\tau$, figure \ref{fig2} displays these stable manifolds for the specific example of $\tau=4$.

A different way of subdividing the state space is provided by the definition of allergic states given in \eqref{eqallergic}. The boundary of the set of allergic states (corresponding to a reference dose) is given by
\begin{align}
\lim_{t\rightarrow\infty}\frac{T_2(t)}{T_r(t)}=\frac{T_2^0}{T_r^0}\,,
\label{eqseparatrix}
\end{align}
that is, by all the states starting from which a single injection does not change the long term ratio $T_2/T_r$. This boundary also forms a two-dimensional manifold in the state space and we will refer to it as the \emph{separatrix}.

\begin{figure}[h!]
\centering
\psfrag{ 1e-4}{\footnotesize$10^{-4}$}
\psfrag{ 0.01}{\footnotesize$0.01$}
\psfrag{ 1}{\footnotesize$1$}
\psfrag{ 100}{\footnotesize$100$}
\psfrag{T1}{\footnotesize$T_1$}
\psfrag{T2}{\footnotesize$T_2$}
\psfrag{Tr}{\footnotesize$T_r$}
\psfrag{T2-Tr-Separatrix}{separatrix}
\psfrag{stabile Mannigfaltigkeit}{stable manifold}
\includegraphics[width=\columnwidth]{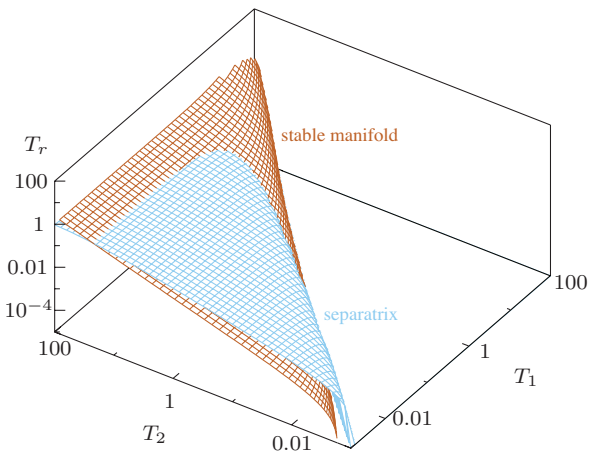}
\caption{Treatable allergic states. Separatrix (light blue) for $D=1$ and one of the stable manifolds (brown) corresponding to $\mathbf{S}_{D,\tau}$ with $\tau=1$. The area between the two surfaces represents the set of allergic states starting from which a successful therapy is possible.}
\label{eqsepstab}
\end{figure}

The goal of specific immunotherapy is to drive the system from an initially allergic state to a tolerant state characterized by increased generation of regulatory T cells. In our model this means approaching the Treg-dominated stable fixed point of the stroboscopic map. Thus, the initial state must lie in the domain of attraction of this fixed point. The crucial question therefore is: \textit{Are there allergic states in the domain of attraction of the ``healthy" fixed point?} In numerical simulations we can show that this is actually the case, provided that the period $\tau$ is not to long. In figure \ref{eqsepstab} the relevant part of the separatrix is shown along with one of the stable manifolds corresponding to $\tau=1$.
The states that lie \emph{below} the separatrix (i.e. in the allergic region) but \emph{above} the stable manifold (i.e. in the domain of attraction of the healthy fixed point) are the ones that allow for a successful therapy. It turns out that this set of treatable allergic states increases if we reduce the period between injections. Also it increases if we choose a higher allergen dose. We can explain this by recalling our investigation of the ratios of cell concentrations. We had found in \eqref{eq:param}, that above a certain threshold the Treg cells will have the highest growth rate independently of the concentrations of the other two cell types. Therefore, even if in comparison there are less Treg cells, they may be able to catch up. Short intervals and high doses result in high concentrations of all T cell subsets and from this especially the Treg cells will profit.
	
\section{Successful Therapy}
\begin{figure}[h!]
\centering
\psfrag{ 1e-04}[l]{\footnotesize$10^{-4}$}
\psfrag{ 0.01}{\footnotesize$0.01$}
\psfrag{ 1}{\footnotesize$1$}
\psfrag{ 100}{\footnotesize$100$}
\psfrag{T}[t]{\footnotesize$T$}
\psfrag{T1}{\scriptsize$T_1$}
\psfrag{T2}{\scriptsize$T_2$}
\psfrag{T3}{\scriptsize$T_r$}
\psfrag{t}{\footnotesize$t$}
\psfrag{ 0}{\footnotesize$0$}
\psfrag{ 5}{\footnotesize$5$}
\psfrag{ 10}{\footnotesize$10$}
\psfrag{ 15}{\footnotesize$15$}
\psfrag{ 20}{\footnotesize$20$}
\psfrag{ 25}{\footnotesize$25$}
\psfrag{ 30}{\footnotesize$30$}
\includegraphics[width=\columnwidth]{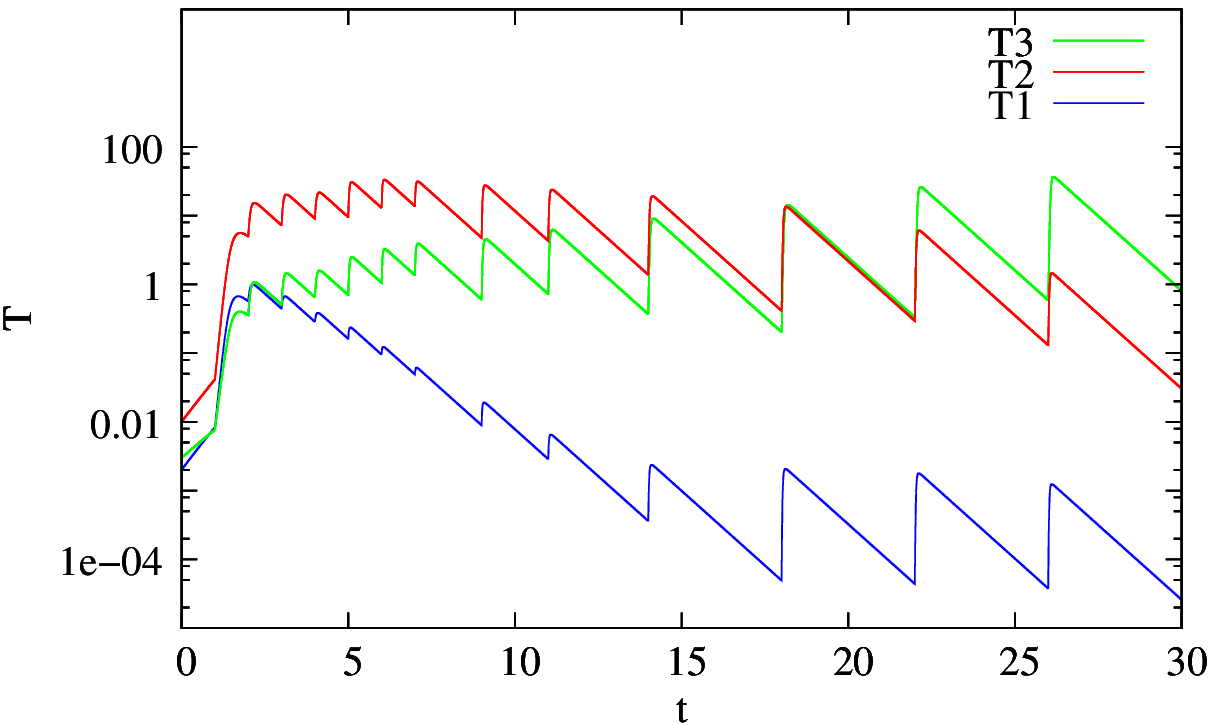}\\
\vspace{0.5cm}
\psfrag{ 1e-4}[l]{\footnotesize$10^{-4}$}
\psfrag{ 0.01}{\footnotesize$0.01$}
\psfrag{ 1}{\footnotesize$1$}
\psfrag{ 100}{\footnotesize$100$}
\psfrag{T}[t]{\footnotesize$T$}
\psfrag{T1}{\scriptsize$T_1$}
\psfrag{T2}{\scriptsize$T_2$}
\psfrag{T3}{\scriptsize$T_r$}
\psfrag{t}{\footnotesize$t$}
\psfrag{ 0}{\footnotesize$0$}
\psfrag{ 5}{\footnotesize$5$}
\psfrag{ 10}{\footnotesize$10$}
\psfrag{ 15}{\footnotesize$15$}
\psfrag{ 20}{\footnotesize$20$}
\psfrag{ 25}{\footnotesize$25$}
\psfrag{ 30}{\footnotesize$30$}
\includegraphics[width=\columnwidth]{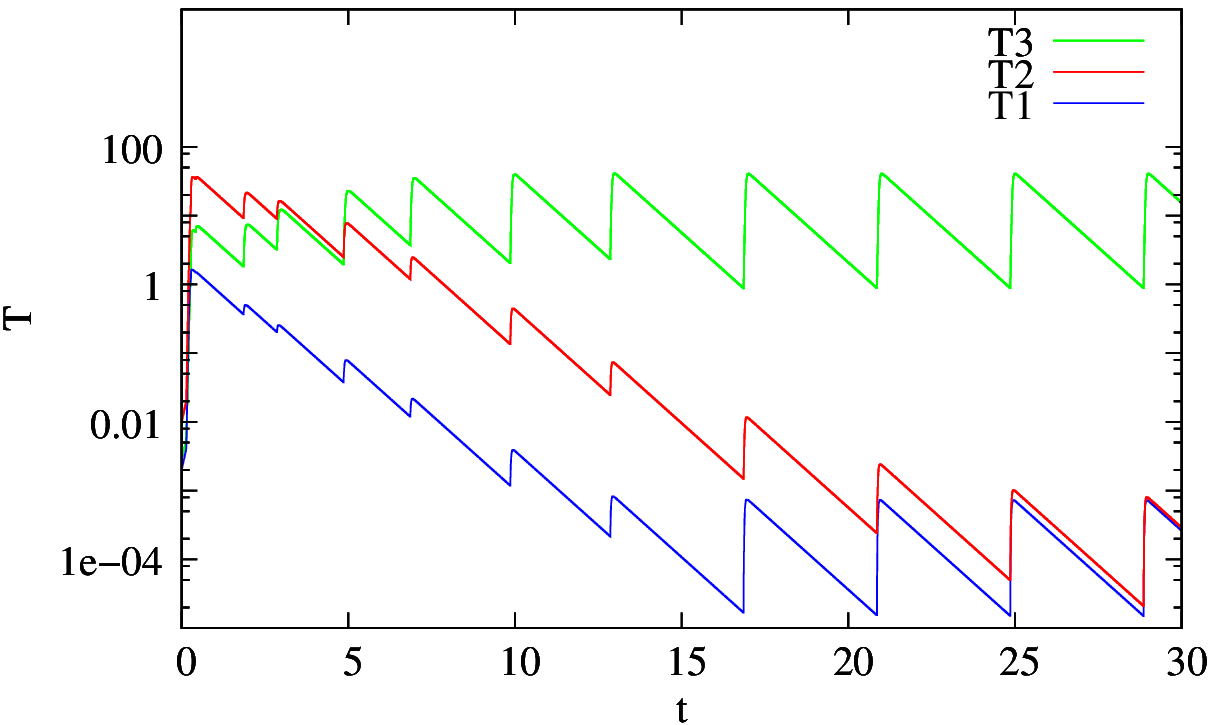}
\caption{Successful therapy. Development of T cell concentrations according to a conventional protocol (top) and rush-protocol (bottom). Initial concentrations in both cases are given by $\mathbf T=(0.002,0.01,0.003)^T$. In the conventional therapy the Treg cells start dominating at about $t=20$, in the rush therapy already at $t=5$. This corresponds to the time when the highest dose is administered for the first time. The therapies are simulated according to protocols found in \cite{rueff2000diagnose} . We assume that the maintenance dose of allergen corresponds to $D=1$ and that one unit of time roughly corresponds to one week.}
\label{figrush}
\end{figure}

	So far we have only looked at the simplified case of periodic therapies, but in computer simulations we can also test protocols as they are used in practice. In these protocols both intervals between injections and administered doses vary. Thus, we cannot directly apply the results that we have achieved in the previous section.  But it is nevertheless still true that high doses and short intervals are in favor of the Treg population. The different protocols that are used in medical practice \citep{rueff2000diagnose} have in common that right before the maintenance phase sets in, the maximum dose is given several times in short intervals. From our point of view this is the decisive step in therapy. It ensures that the system reaches the domain of attraction of the stable fixed point that the therapy is aiming at. At this stage the Treg cells start overriding the Th2 cells. In figure \ref{figrush} simulations according to a conventional protocol and according to a rush-protocol are shown.
In the conventional protocol the maintenance phase sets in after about two months, in the rush protocol after only a few days. Consequently the therapeutic effect sets in much earlier in rush-protocols, which is in accordance to what has been observed in practice \citep{cox2008advantages}. In both protocols the same periodic orbit is finally reached because the maintenance phase is the same for both therapies. Therefore, the final result of therapy is the same and independent of the protocol. 

In our simulations we also find that the Th1/Th2 ratio increases during therapy. At the beginning $T_2$ is clearly higher than $T_1$, whereas at the end both cell concentrations have the same order of magnitude. This might be an explanation for the ``Th2-Th1 switch'' that has been observed during specific immunotherapy.
	
\section{Conclusion}
	Our model is able to describe allergic reactions and the course and outcome of allergen-specific immunotherapy on the T cell level. Apart from this work and the models that it is directly based on, we find other attempts in the literature to explain specific immunotherapy by means of mathematical descriptions of T cell dynamics, for example \cite{fishman1996modeling}. To our knowledge, however, the present paper is the first one to include regulatory T cells. We have shown that the basic mechanisms in allergic reactions can be explained as a competition between Th2 and Treg cells. In the model, Treg responses are favored by high allergen doses administered in short time intervals. Therefore, the decisive event in immunotherapy is the beginning of the maintenance phase. Protocols in practice could be improved by optimizing this step. As immunologists have not yet fully understood all the regulatory mechanisms playing a role in allergic diseases, we are sure that our model will need further refinement. Nevertheless, our investigations already provide general tools to model immune reactions with interacting lymphocytes. Once the immunological picture of T cell regulation is more complete, we will be able to give a more adequate description of the real system. In particular, it would be interesting to extend the model further by including recently identified T cell subsets, such as Th17 cells, which are found to play a role in allergic asthma \citep{oboki2008th17,schmidt2007th17}.

\bibliography{references}

\end{document}